\documentclass{aa}  
\usepackage{aalongtable,lscape}
\usepackage{graphicx}
\usepackage{txfonts}
\begin{document}
\title{Reaching the boundary between \\ stellar kinematic groups and very wide
binaries. II}  
\titlerunning{$\alpha$~Lib + KU~Lib}
\subtitle{$\alpha$~Lib + KU~Lib: a common proper motion system in Castor
separated by 1.0\,pc}
\author{Jos\'e A. Caballero\inst{1,2}}
%
%
\institute{
Departamento de Astrof\'{\i}sica y Ciencias de la Atm\'osfera, Facultad de
F\'{\i}sica, Universidad Complutense de Madrid, 28040 Madrid, Spain,
\email{caballero@astrax.fis.ucm.es} 
\and
Centro de Astrobiolog\'{\i}a (CSIC-INTA), Carretera de Ajalvir km~4,
28850 Torrej\'on de Ardoz, Madrid, Spain}
\date{Received 30 December 2009; accepted 21 January 2010}

\abstract
{} 
{I investigate the gravitational binding of a nearby common proper motion system
in the young Castor moving group ($\tau \sim$ 200\,Ma), which is formed by the
bright quadruple star $\alpha$~Lib (Zubenelgenubi) and the young solar analogue
KU~Lib. 
The system has an exceptionally wide angular separation of about 2.6\,deg,
which corresponds to a projected physical separation of about 1.0\,pc.} 
{I compiled basic information on the system, compared its binding energy with
those of other weakly bound systems in the field, and studied the 
physical separations of resolved multiple systems in Castor.}   
{KU~Lib has roughly the same proper motion, parallactic distance, radial
velocity, and metallicity as the young hierarchical quadruple system
$\alpha$~Lib.
It also displays youth features.
The resemblance between these basic parameters and the relatively high
estimated binding energy indicate that the five stars are gravitationally
bound. 
KU~Lib and $\alpha$~Lib constitute the widest known multiple system in all mass
domains, and probably represent the most extreme example of young wide binaries
on the point of being disrupted.
Besides this, I~make a comprehensive compilation of star candidates in Castor,
including new ones.}  
{}
\maketitle
%

\section{Introduction}
\label{introduction}

What is the widest separation between stars in a binary system?
To answer this question, we must enter a range of separations in which
arseconds and astronomical units are almost useless and must instead use 
arcminutes (or degrees) and thousands of astronomical units (or parsecs).
For example, the nearest star to the Sun, \object{Proxima~Centauri}, is located
at an angular separation of about 185\,arcmin (3.1\,deg) to the binary system
\object{$\alpha$~Cen~A} and~B, which at the heliocentric distances of the
components is translated into a physical separation $r$ = 12.0$\pm$0.6\,kAU
(0.058$\pm$0.003\,pc -- Innes 1915; Wertheimer \& Laughlin 2006). 
Although \"Opik (1932) had theoretically considered maximum apoapsides of
up to 200\,kAU ($\sim$1\,pc) in his novel investigation of the gravitational
disruption of wide binary stars, atferwards consensus developed in the
literature that there is a cutoff at $s \sim$ 20\,kAU ($\sim$0.1\,pc) in
projected physical separation of binaries (e.g. Tolbert 1964; Heggie 1975;
Bahcall \& Soneira 1981; Retterer \& King 1982; Weinberg et~al. 1987; Gliese \&
Jahreiss 1988; Close et~al. 1990 -- but see Quinn et~al. 2009).  
However, state-of-the-art analytical studies, such as the ones by Jiang \&
Tremaine (2009), allow the existence of wide binary stars separated by 200\,kAU
or more. 
Some of these systems may be formerly bound binaries that are slowly
drifting~away. 
 
Multiple systems in the solar neighbourhood with $s >$ 20\,kAU do exist.
Six of them are Washington double stars in the astro-photometric follow-up of
Caballero (2009), of which \object{HD~6101}~AB and \object{G~1--45}~AB ($s$ =
26.9$\pm$0.6\,kAU), \object{$\xi^{02}$~Cap} and \object{LP~754--50} ($s$ =
28.3$\pm$0.3\,kAU), and \object{HD~200077}~AE--D and \object{G~210--44}~AB ($s$
= 49.7$\pm$1.1\,kAU) had higher binding energies (in absolute value) than
$\alpha$~Cen~AB and Proxima (see Sect.~\ref{bindingenergyand}). 
Other systems with even wider projected physical separations of more than
50\,kAU (a quarter parsec) have been reported.  
Some of the very wide multiple systems in Table~\ref{table.verywidebinaries}
belong to the Galactic halo, where the probability of encountering stars is at a
minimum (e.g. Quinn \& Smith 2009), or to young moving groups, such as the
\object{Hyades supercluster} or the \object{Castor moving group}. 
Young stars have had ``less time to encounter individual stars and giant
molecular clouds, whose gravity will eventually tear them apart'' (Caballero
2009 and references therein).
Other wide systems have been proposed by Wasserman \& Weinberg (1991),
Allen et~al. (2000), or Chanam\'e \& Gould (2004), but many of them do not pass
a simple common proper motion filter with the Aladin sky atlas (Bonnarel et~al.
2000) and USNO-B1 data (Monet et~al. 2003).

   \begin{table}
      \caption[]{Very wide binaries and candidates in the literature with
      projected physical separations $s >$ 50\,kAU.} 
         \label{table.verywidebinaries}
     $$ 
         \begin{tabular}{l l c l}
            \hline
            \hline
            \noalign{\smallskip}
Primary  		& Secondary		& $s$      	& Reference$^{a}$ \\
  			& 			& [kAU]		& 		\\
            \noalign{\smallskip}
            \hline
            \noalign{\smallskip}
\object{LP~268--35}$^{b}$	& \object{LP~268--33}	& 84$\pm$15 	& LB07, Ca09	\\ %
\object{HD~136654}$^{c}$	& \object{BD+32~2572}	& 68.8$\pm$1.7 	& LB07, Ca09	\\ %
\object{HD~149414}~AB$^{d}$	& \object{BD--05~3968B}	& $\sim$55.7	& ZOM04		\\ %
\object{Ross 570}$^{d}$		& \object{WT 1370}	& $\sim$55.4	& Al00		\\ %
\object{V869~Mon}~AB		& \object{BD--02~2198}	& $\sim$55.3	& Po09		\\ %
\object{Fomalhaut}$^{e}$	& \object{TW~PsA}	& $\sim$54.4	& Gl69		\\ %
\object{BD--12~6174}		& \object{LP~759--25}	& $\sim$50.5	& Ma08		\\ %
           \noalign{\smallskip}
            \hline
         \end{tabular}
     $$ 
\begin{list}{}{}
\item[$^{a}$] References --  
Gl69: Gliese 1969;
Al00: Allen et al. 2000;
ZOM04: Zapatero Osorio \& Mart\'{\i}n 2004;
LB07: L\'epine \& Bongiorno 2007;
Ma08: Makarov et al. 2008;
Po09: Poveda et~al. 2009;
Ca09: Caballero 2009.
\item[$^{b}$] Binary candidate of unknown true status.
\item[$^{c}$] Young binary in the Hyades Supercluster.
\item[$^{d}$] Halo ``metal-poor'' binaries with large distance uncertainties.
\item[$^{e}$] Young binary in the Castor moving group.
\end{list}
   \end{table}

This analysis is a continuation of the virtual observatory study in
Caballero (2009), where I~started a programme of identifying and investigating
the widest common proper motion pairs.

\section{The $\alpha$~Lib + KU~Lib system}
\label{thesystem}


During a search for pairs of {\em Hipparcos} stars with small relative
differences in proper motions and spatial vectors\footnote{The spatial vector of
a star was defined by $\vec{r} = (x,y,z) \equiv (d \cos{b} \cos{l}, d \cos{b}
\sin{l}, d \sin{b}$), where $d$ is the heliocentric distance of the star and $l$
and $b$ are the Galactic longitude and latitude, with the suitable sign
convention.} but large angular separations (Caballero, in~prep.), I~noticed the
resemblance between the parallaxes and proper motions of two stars separated by
155.8\,arcmin (about 2.6\,deg); see Fig.~\ref{theimage}. 
A~quick glance at the system with Aladin and SIMBAD showed that the primary is
actually the brightest star of a known hierarchical quadruple system
(Table~\ref{table.thesystem}).
With $V$ = 2.75\,mag, the primary $\alpha^{02}$~Lib~AB (Zubenelgenubi, {\em
az-Zuban al-Janubi}, ``the southern claw''), is the second brightest star in the
constellation of Libra after \object{$\beta$~Lib} (Zubeneschamali, {\em az-Zuban
ash-Shamali}, ``the northern claw'').   
The star $\alpha^{02}$~Lib~AB merited not only an Arabic name and a Bayer
designation, but also a Chinese name of a zodiacal star in the 4--6th centuries
(Liu 1986). 
The spectroscopic binarity of $\alpha^{02}$~Lib~AB, with velocity variations of
at least 80\,km\,s$^{-1}$ in about 60\,d, have already been noticed by Slipher
(1904).
Additional variable radial velocity measurements have been puplished by a number
of authors (e.g. Lee 1914; Young 1917; Wilson 1953).
While most of the spectral type determinations of the brightest component are
consistent with a peculiar A3--4 dwarf-subgiant classification, there is
no consensus on the spectral type of the faintest component, which is not
visible in the spectra.
Besides, the system has never been resolved by speckle interferometry or any
other imaging technique (e.g. Hartkopf \& McAlister 1984).

Based on observations taken in June 1823, Herschel \& South (1824) first
tabulated a companion of the ``6th magnitude'' to the northwest of
$\alpha^{02}$~Lib~AB. 
They tabulated (with modern notation) an angular separation $\rho$ =
230.853\,arcsec and a position angle $\theta$ = 315.45\,deg.
Almost two centuries later, these values have stayed constant within
uncertainties at $\rho$ = 230.9$\pm$0.3\,arcsec, $\theta$ =
316.01$\pm$0.11\,deg, in spite of the system having travelled together more than
20\,arcsec during this time.
The proper-motion companion at about 5.4\,kAU to $\alpha^{02}$~Lib~AB is
$\alpha^{01}$~Lib~AB, which is in turn a single-lined spectroscopic
binary with  $\gamma$ = --23.47$\pm$0.15\,km\,s$^{-1}$ and K$_1$ =
3.69$\pm$0.17\,km\,s$^{-1}$ (Duquennoy \& Mayor 1991).
Using near-infrared adaptive optics, $\alpha^{01}$~Lib~AB was later resolved in
a 0.383\,arcsec-wide pair ($s \sim$ 0.01\,kAU) with $\Delta H \approx$ 3.40\,mag
by Beuzit et~al. (2004), who also provided a definitive orbital period of $P
\approx$ 5870\,d based on radial-velocity measurements.
The system offers good prospects for determining reasonably accurate masses.
Afterwards, Makarov \& Kaplan (2005) catalogued it as an astrometric binary
with accelerating proper motion in {\em Hipparcos}\footnote{Besides, Pannunzio
et~al. (1992) confused $\alpha^{01}$~Lib~AB with a fainter star ($V \sim$
13.2\,mag) at 1.8\,arcmin to the southwest, which they called \object{AOT~53}.
However, the proper motion tabulated in the Positions and Proper Motions
eXtended catalogue (PPMX -- R\"oser et~al. 2008) is inconsistent with membership
in the $\alpha$~Lib system.}. 

The Vega-like status of the primary $\alpha^{02}$~Lib~AB (see below) and
the X-ray activity of the secondary $\alpha^{01}$~Lib~AB (Morale et~al. 1996;
H\"unsch et~al. 1998) led several authors to classify them as a nearby young
system, with an age of less than 800\,Ma (e.g. Duncan 1984; Artymowicz \&
Clampin 1997; Rieke et~al. 2005). 
Based on kinematics criteria, the quadruple system was listed as a member in the
Castor moving group by Barrado y Navascu\'es (1998) and Montes et~al. (2001a). 
Subsequently, Ribas (2003) cast doubts on the membership in the moving group
of the A-type star $\alpha^{02}$~Lib~AB based on a faint overluminosity with
respect to theoretical isochrones in a colour-magnitude diagram, but he did not
account for the known binarity.
The latest determinations of the most probable age of the Castor moving group
point towards about 200\,Ma (e.g. L\'opez-Santiago et~al. 2009)\footnote{It
is still under discussion whether the Castor moving group is younger
(80--200\,Ma) but contaminated by older stars of Hyades-like age of the
\object{Centaurus-Crux} association and the \object{Coma} cluster (Chereul
et~al. 1999; Asiain et~al. 1999; L\'opez-Santiago 2005).}. 
Because of its youth, the quadruple system has also been the target of searches
for circumstellar discs in the visible (Smith et~al. 1992; Holweger et~al.
1999), infrared (Aumann \& Probst 1991; Oudmaijer et~al. 1992; Cheng et~al.
1992; Rieke et~al. 2005), and microwave (Slee \& Budding 1995).  
Known to be associated to the infrared source \object{IRAS~14479--1547}, the
F-type star $\alpha^{01}$~Lib~AB has strong flux excesses at 12, 70, and,
especially, 24\,$\mu$m due to a dusty disc and, possibly, mid-infrared emission
features. 
The grains around $\alpha^{01}$~Lib~AB are probably very warm, and the material
around the star must be replensihed from a reservoir (Chen et~al. 2005). 

   \begin{table}
      \caption[]{The $\alpha$~Lib + KU~Lib system$^{a}$.} 
         \label{table.thesystem}
     $$ 
         \begin{tabular}{l cc cc cc l c}
            \hline
            \hline
            \noalign{\smallskip}
Star  					& 
\object{$\alpha^{02}$ Lib} AB	& 
\object{$\alpha^{01}$ Lib} AB	& 
\object{KU Lib}		\\
            \noalign{\smallskip}
            \hline
            \noalign{\smallskip}
Flamsteed				& 9 Lib			& 8 Lib			& ...			\\
HD					& 130841		& 130819		& 128987		\\
HIP					& 72622			& 72603			& 71743			\\
$\alpha^{J2000}$			& 14 50 52.71		& 14 50 41.18		& 14 40 31.11 		\\      
$\delta^{J2000}$      			& --16 02 30.4   	& --15 59 50.0		& --16 12 33.4  	\\   
$B$ [mag]				& 2.91			& 5.52			& 7.92			\\
$V$ [mag]				& 2.75			& 5.15			& 7.24			\\
Spectral type				& A4\,IV--V + F:	& F4\,V + M:		& G8\,V\,(k)	 	\\	
$\mu_\alpha \cos{\delta}$ [mas\,a$^{-1}$]& --105.7$\pm$0.2  	& --136.3$\pm$0.4  	& --112.0$\pm$0.5 	\\				
$\mu_\delta$ [mas\,a$^{-1}$]  		& --68.40$\pm$0.12	& --59.0$\pm$0.3	& --65.0$\pm$0.4 	\\		
$d$ [pc]				& 23.24$\pm$0.10	& 23.0$\pm$0.2		& 23.7$\pm$0.3  	\\ 		
$V_r$ [km\,s$^{-1}$] 			& +20/--60		& --23.47$\pm$0.15 	& --23.3$\pm$0.2 	\\		
U [km\,s$^{-1}$] 			& ...		 	& --25		 	& --23		 	\\		
V [km\,s$^{-1}$] 			& ...		 	& --8		 	& --7		 	\\		
W [km\,s$^{-1}$] 			& ...		 	& --13		 	& --14		 	\\		
{[Fe/H]} 				& ...		 	& --0.07		& +0.02		 	\\		
           \noalign{\smallskip}
            \hline
         \end{tabular}
     $$ 
\begin{list}{}{}
\item[$^{a}$] Coordinates J2000, proper motions, and parallactic distances from
van~Leeuwen (2007), $B$ and $V$ magnitudes from Perryman et~al. (1997), spectral
types of primaries from Gray et~al. (2006), $UVW$ space velocities and [Fe/H]
metallicities from Holmberg et~al. (2009), and radial velocities from Slipher
(1904 --  $\alpha^{02}$~Lib~AB), Duquennoy \& Mayor (1991 --
$\alpha^{01}$~Lib~AB), and Nordstr\"om et~al. (2004 -- KU~Lib).
\end{list}
   \end{table}
%


The ``secondary'' in my {\em Hipparcos} search at 2.6\,deg to $\alpha$~Lib was
the young solar analogue KU~Lib.
With proper motion, parallactic distance, radial velocity, and metallicity
similar to those of the $\alpha$~Lib system (Table~\ref{table.thesystem}),
KU~Lib also displays youth features: it is a prominent X-ray
emitter (Gaidos 1998), 
shows astrospheric absorption of the stellar H~{\sc i} Ly$\alpha$ emission line
(in spite of the low interstellar-medium wind velocity seen by the star -- Wood
et~al. 2005a, 2005b), and has a high lithium abundance $\log{\epsilon(\rm{Li})}$
and a short period of photometric variability of $P$ = 9.35$\pm$0.04\,d,
possibly associated to fast rotation and photospheric dark spots (Gaidos et~al.
2000). 
Plavchan et~al. (2009) gave KU~Lib a Hyades-like age (i.e. $\tau \sim$ 
600\,Ma)
based on  X-ray--age and rotation--age correlations.
There have been numerous determinations of effective temperature, spectral type,
and metallicity for KU~Lib, consistent with late-G dwarf and solar abundance
(Feltzing \& Gustafsson 1998; Haywood 2001; Gaidos \& Gonz\'alez 2002; Valdes
et~al. 2004; Nordstr\"om et~al. 2004).
KU~Lib has no flux excesses in the 24, 30--34, and 70\,$\mu$m {\em 
Spitzer} bands (Lawler et~al. 2009; Plavchan et~al. 2009).
Finally, the star is an exoplanet target with no known planetary companion
(Grether \& Lineweaver 2006). 


\section{Discussion}

The stars $\alpha$~Lib and KU~Lib are located at the same distance, move in
exactly the same direction in space, and probably have the same age and
metallicity (hence formed from the same parental cloud).
However, they are separated by an enormous projected physical separation of $s
\approx$ 217\,kAU ($\sim$1.0\,pc).
This value nearly quadruples the separations of the most reliable wide
binaries in Table~\ref{table.verywidebinaries} and is close to the
typical separation between stars in the Galactic disc (e.g. Proxima is located
at $d$ = 1.296$\pm$0.004\,pc to the Sun). 
An inquisitive reader may consider that common distance, proper motion,
age, metallicity, and location in the sky are disputable evidence of
$\alpha$~Lib and KU~Lib constituting the widest known multiple system (see e.g.
Scholz et~al. 2008 for the example of a debatable, extremely wide, very low-mass
pair with common proper motion).  
Besides, the proper motions of $\alpha^{02}$~Lib~AB, $\alpha^{01}$~Lib~AB,
and KU~Lib are not identical, and there is no determination of the
systemic radial velocity, $\gamma$, of the A-type primary.
However, according to Beuzit et~al. (2004), the mildly discrepant {\em
Hipparcos} proper motions of $\alpha^{02}$~Lib and $\alpha^{01}$~Lib are
consistent with the orbital motion of the photocentre of the secondary over the
{\em Hipparcos} mission. 
The $\alpha$~Cen and Proxima system is the most obvious example that wide
physically bound systems can still have slightly different proper motions 
(Wertheimer \& Laughlin 2006; Caballero 2009). 
Lastly, the radial velocity measurements of $\alpha^{02}$~Lib~A(B) are centred
on the average value $\gamma \sim$ --20\,km\,s$^{-1}$, consistent with the ones
of $\alpha^{01}$~Lib~AB and KU~Lib.

\subsection{Resolved multiple systems in Castor}

There are other young stars in moving groups that form wide (and ver wide)
multiple systems. 
The stars \object{HD~136654} and \object{BD+32~2572} in the {Hyades
Supercluster} (Table~\ref{table.verywidebinaries}) and \object{AU~Mic} and
\object{AT~Mic~A}B in the \object{$\beta$~Pictoris moving group}, with a
projected physical separation $s$ = 46.4$\pm$0.5\,kAU, are among the widest 
young multiple systems (Caballero 2009).

   \begin{table}
      \caption[]{Resolved multiple system candidates in the Castor moving group.} 
         \label{table.castormultiple}
     $$ 
         \begin{tabular}{ll ccc c}
            \hline
            \hline
            \noalign{\smallskip}
Primary  		& Secondary		& $\rho$	& Ref.$^{a}$ 	& s 		\\
  			& 			& [arcsec]	& 		& [kAU]		\\
            \noalign{\smallskip}
            \hline
            \noalign{\smallskip}
V450~And~AB  		& V451~And		& 615.1		& TYC		& 16.3		\\ %
V575~Pup~A  		& V575~Pup~B		& 2.587		& HIP		& 0.078		\\ 
V356~CMa~A  		& V356~CMa~B		& 2.010		& FM00		& 0.10		\\ 
HD~51825~A  		& HD~51825~B		& 0.140		& HIP		& 0.006		\\ 
VV~Lyn~AB  		& BL~Lyn		& 33.88		& TYC		& 0.40 		\\ 
Castor~A  		& Castor~B		& 3.124		& HIP		& 0.049		\\ 
Castor~AB  		& YY~Gem		& 70.81		& TYC		& 1.10		\\ 
CU~Cnc~AB  		& CV~Cnc~AB  		& 10.16		& 2M		& 0.11		\\ 
$\psi$~Vel~A  		& $\psi$~Vel~B		& 0.679		& HIP		& 0.013		\\ 
36~UMa~A  		& 36~UMa~BC		& 122.8		& TYC		& 1.57		\\ 
HD~119124~A  		& HD~119124~B		& 3.979		& 2M		& 0.10		\\ %
$\alpha^{01}$ Lib A	& $\alpha^{01}$ Lib B 	& 0.383		& Be04		& 0.009		\\ 
$\alpha^{02}$~Lib~AB 	& KU~Lib		& 9344		& 2M		& 217		\\ 
$\alpha^{02}$~Lib~AB 	& $\alpha^{01}$ Lib AB 	& 231.0		& 2M		& 5.37		\\ 
$\mu$ Dra A		& $\mu$ Dra B 		& 2.160		& HIP		& 0.059		\\ %
$\mu$ Dra AB		& $\mu$ Dra C 		& 12.45		& 2M		& 0.34		\\ %
V1436~Aql~A  		& V1436~Aql~B		& 17.33		& 2M		& 0.44		\\ 
HD~186922~A  		& HD~186922~B		& 0.159		& HIP		& 0.005		\\ 
V447~Lac~A  		& V447~Lac~B		& 76.79		& 2M		& 1.65		\\ %
Fomalhaut  		& TW~PsA		& 7064		& 2M		& 54.4		\\ 
HD~218739~AB  		& KZ~And~AB		& 15.64		& TYC		& 0.38		\\ 
EQ~Peg~A  		& EQ~Peg~B		& 5.029		& HIP		& 0.031		\\ 
           \noalign{\smallskip}
            \hline
         \end{tabular}
     $$ 
\begin{list}{}{}
\item[$^{a}$] References --
HIP: {\em Hipparcos}, Perryman et~al. (1997);
TYC: Tycho-2, H{\o}g et~al. (2000);
FM00: Fabricius \& Makarov (2000);
Be04: Beuzit et~al. (2004);
2M: 2MASS, Skrutskie et~al. (2006).
\end{list}
   \end{table}

There might be other wide systems in the Castor moving group, with 
separations of the order of 50--200\,kAU.
To determine the typical and maximum separations between multiple systems in
Castor and to quantify the exceptionality of the $\alpha$~Lib and KU~Lib system,
I~first had to compile a comprehensive list of candidate members, which is shown
in Table~\ref{table.castor}. 
To assemble it, I~merged the shorter lists published by Barrado y Navascu\'es
(1998), Montes et~al. (2001a, 2001b), Ribas (2003), and L\'opez-Santiago et~al.
(2010). 
I~discarded nine spectroscopic binaries\footnote{The discarded distant
spectroscopic binaries were: 5~Cet, HD~43516, VV~Mon, FI~Cnc, VX~Pyx, FF~UMa, 
EQ~Leo, IL~Com, and V894~Her.} with {\em Hipparcos} parallactic distances $d 
\gtrsim$ 100\,pc and six stars that have been afterwards ascribed to other 
moving groups (AG~Tri~AB and AU~Mic to the $\beta$~Pictoris moving group; 
HD~98736~AB, BD+08~2599, and HD~162283 to the \object{Local Association}) or to 
the field ($\iota$~Peg~AB). 
This made a list of 70 star candidates in the Castor moving group.
Coordinates were taken from the {\em Hipparcos} catalogue except for a few faint
targets, which were taken from the Tycho-2 or 2MASS catalogues, depending on
availability. 
Heliocentric distances were retrieved from van~Leeuwen (2007), except for 
\object{LP~944--20} (from Tinney 1996),
\object{DX~Cnc} and \object{V1436~Aql}~AB (from van~Altena et~al. 1995), 
and \object{AD~Leo} (from Jenkins 1952). 
The distance to \object{QT~And} is not parallactic, but photometric, and was 
estimated by Carpenter et~al. (2005). 

In the list in Table~\ref{table.castor}, apart from KU~Lib, there are two
known common proper motion secondaries and two primaries that had never been
listed as members in the Castor moving group.
They are 
\object{VV~Lyn}~AB (M2.5; companion to BL~Lyn),
\object{36~UMa~A} (F8V),
\object{$\mu$~Dra~C} (M3; also known as GJ~9584~C),
and \object{V447~Lac~B} (M4; G~232--62).
Besides these, three stars, \object{BD--46~11540} (Johnson et~al. 2007),
\object{Fomalhaut} (Kalas et~al. 2008; in imaging), 
and \object{HD~217107} (Marcy et~al. 1999; Vogt et~al. 2005), 
have exoplanet candidates, which supports the Caballero et~al. (2009) hypothesis
that some young stars ($\tau \lesssim$ 600\,Ma) pass the activity filters for
radial velocity searches (see also Paulson \& Yelda 2006).

A large fraction of the Castor star candidates in Table~\ref{table.castor} show
some kind of multiplicity: 15 (21\,\%) of them are spectroscopic binaries yet
unresolved by imaging, nine (13\,\%) are close binaries only resolved by {\em
Hipparcos} and, in some cases, adaptive optics or micrometer techniques, and 13
(19\,\%) are wide binaries resolved in the Tycho-2 and/or 2MASS catalogues.
Accounting for all types of multiplicity (there are a few hierarchical systems
containing wide and spectroscopic or close binaries), 40 stars (57\,\%) are part
of multiple systems.

Table~\ref{table.castormultiple} provides the angular and projected physical
separations of the 22 resolved multiple systems in Castor.
All the systems were known except for the ``pairs'' $\alpha$~Lib and KU~Lib, and
V450~And~AB and V451~And.  
Within uncertainties, the two latter solar-like stars have the same parallaxes
and proper motions as in {\em Hipparcos}, radial velocities and metallicities
as in Nordstr\"om et~al. (2004) and Karata{\c s} et~al. (2005), and gyrochronous
and kinematic ages as in Strassemier et~al. (2000) and Montes et~al. (2001a). 
The primary is a single-lined spectroscopic binary with acceleration of proper
motion that was suspected for some time to harbour an exoplanet (Perrier et~al.
2003; Makarov \& Kaplan 2005). 
V450~And~AB and V451~And are separated by about 16\,kAU.
With individual masses of about 1.1\,$M_\odot$, the system is likely to be bound
(see below).

Of the 22 resolved multiple systems, only seven have projected physical
separations $s >$ 1\,kAU. 
Three of them, Castor~AB and YY~Gem, 36~UMa~A--BC, and V447~Lac~AB, have
relatively small separations of $s \approx$ 1.10--1.65\,kAU.
The four other wide ``pairs'' are the new triple V450~And~AB and V451~And, the
Gliese (1969)'s binary Fomalhaut (A4V) and TW~PsA (K4V), with $s \approx$
54.4\,kAU, and the $\alpha$~Lib and KU~Lib system ($\alpha^{01}$~Lib and 
KU~Lib are located at about 5.37 and 217\,kAU to $\alpha^{02}$~Lib,
respectively).

Finally, I~investigated the separations from the closest neighbours of the stars
in Table~\ref{table.castor} after discarding the multiple systems in
Table~\ref{table.castormultiple}, and found that the minimum angular separation 
was between Castor~AB and VV~Lyn~AB, of $\rho \sim$ 4.4\,deg.  
Although this angular separation is only about 1.7 times more than the one
between $\alpha$~Lib and KU~Lib, Castor~AB (and its companion YY~Gem) and
VV~Lyn~AB (and its companion BL~Lyn) are at different heliocentric distances by
almost 4\,pc.
Remaining separations are much larger than this value.
Thus, $\alpha$~Lib and KU~Lib resembles a very wide physical system more than a
fortuitous rendezvous of stars in the Castor moving group.

\subsection{Binding energy and differential Galactic tidal force}
\label{bindingenergyand}

\begin{figure}
\centering
\includegraphics[width=0.49\textwidth]{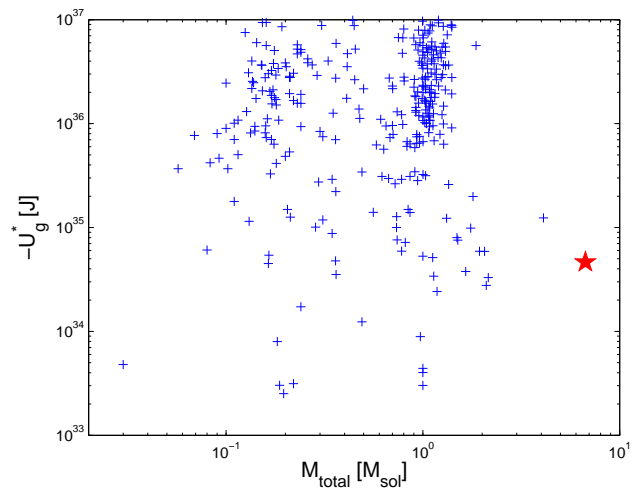}
\caption{Binding energy-total mass diagram.
Crosses are for the 406 systems with very low-mass components and
Washington double stars with the widest angular separations presented in
Caballero (2009).
The system $\alpha$~Lib and KU~Lib is represented by a filled star.
The nearby data point at $U_g^*$ = --120~10$^{33}$\,J and $\mathcal{M}_{\rm
total}$ = 4.1\,$M_\odot$ corresponds to the system HD~200077~AE--D and
G~210--44~AB.}   
\label{thefigure}
\end{figure}

One way to assert physical bounding of the $\alpha$~Lib and KU~Lib system 
is by comparing the absolute value of its binding energy (actually, the minimum 
absolute value $|U_g^*|$ computed using the projected physical separation $s$ instead of
the  true physical separation $r$ -- Caballero 2009) with those of other 
systems of very low-mass or very wide separations.  
For that, it is necessary to estimate the individual masses of the five
components. 
Using the Siess et~al. (2000) models for an age of 200\,Ma, the spectral type
and visual absolute magnitude of $\alpha^{01}$~Lib~AB (derived from $V$ and
$d$), and the $H$-band magnitude difference between components measured by
Beuzit et~al. (2004), I~estimate masses of about 1.4--1.5 and
0.5--0.6\,$M_\odot$ for its A and B components.
The expected spectral type of $\alpha^{01}$~Lib~B is M0--1\,V.
In the same way, $\alpha^{02}$~Lib~A and KU~Lib have estimated masses of 2.2 and
1.0\,$M_\odot$, respectively. 
There is not enough information for an accurate assessment of the mass of
$\alpha^{02}$~Lib~B.
Since it produces a large radial-velocity amplitude on $\alpha^{02}$~Lib~A and
overluminosity in the colour-magnitude diagram, I~guess a spectral type F and a
mass of 1.5\,$M_\odot$; therefore the total estimated mass of the $\alpha$~Lib
and KU~Lib system is 6.7\,$M_\odot$. 
The 1.0\,$M_\odot$-mass star KU~Lib feels the gravitational pull of an
equivalent single star of 5.7\,$M_\odot$.  

The derived binding energy and orbital period of $\alpha$~Lib and KU~Lib are
$U_g^*$ = --46~10$^{33}$\,J and $P \approx$ 39\,Ma.
Although the system may be slowly drifting away and will eventually be
disrupted, it has a value of $|U_g^*|$ that is more than many known binaries
(Fig.~\ref{thefigure}).
For comparison, the $\alpha$~Cen and Proxima system has a true binding energy of
$U_g$ = --32.1~10$^{33}$\,J. 
Furthermore, the $\alpha$~Lib and KU~Lib value of $|U_g^*|$ is one order of
magnitude higher than those of the most fragile systems known: the
Koenigstul\,1-like, very low-mass ($\mathcal{M} <$ 0.2\,$M_\odot$), very wide
($\rho >$ 1\,kAU) binaries (Caballero 2007; Artigau et~al. 2007; Radigan et~al.
2009), and the very young substellar wide binaries, such as the brown-dwarf
exoplanet pair 2M1207--39~AB (Chauvin et~al. 2004 -- see Caballero 2009 for
more examples {\em without} common proper motion confirmation). 
Besides, the orbit of KU~Lib around $\alpha$~Lib may be highly eccentric, so the
value of $|U_g^*|$ could be even higher than computed (and the orbital period
would be shorter). 

According to Close et~al. (1990), ``the physical limit to the maximum
separation [that] a candidate system could possibly have is the separation at
which the differential Galactic force exceeds the gravitational binding force of
the system''.
They computed that the physical limit for a wide binary system of two
1.0\,$M_\odot$-mass stars was $r \sim$ 1.05\,pc.
Since the gravitational binding force is proportional to the product of masses, 
the separation at which the system $\alpha$~Lib and KU~Lib will be torn apart
would be 5.7 (5.7\,$M_\odot$ $\times$ 1.0\,$M_\odot$) times larger, at about
6.0\,pc. 
Even accounting for a generous projection factor for transforming the measured
projected physical separation $s \sim$ 1.0\,pc into a real physical separation
$r$, the computations above allow the existence of $\alpha$~Lib and KU~Lib
during a few revolutions around the Galaxy centre.

\section{Summary}

Because of the resemblance between their parallaxes, proper motions, radial
velocities, metallicities, and most probable ages, I~first propose that
$\alpha^{02}$~Lib~AB, $\alpha^{01}$~Lib~AB, and KU~Lib form a hierarchical
quintuple system separated by about 2.6\,deg.
At the distance of the system, $d \sim$ 23\,pc, the resulting projected
physical separation is $s \sim$ 217\,kAU ($\sim$1.0\,pc), which makes
$\alpha$~Lib and KU~Lib the widest multiple system candidate in all mass
domains. 
The stars are likely members of the young Castor moving group ($\tau \sim$
200\,Ma) and do not seem to be the result of a fortuitous approximation or
alignment within the moving group.
The stars have a binding energy that is greater than many other low-mass
multiple systems reported in the literature and that is consistent with the
system's survival to the Galactic tidal force. 
As spin-offs of this work, I~compiled an exhaustive list of star candidates in
Castor, including five new stars (KU~Lib itself, VV~Lyn~AB, 36~UMa~A,
$\mu$~Dra~C, and V447~Lac~B), enumerated two young stars with radial-velocity
exoplanet candidates, and found a new wide Castor pair separated by about
16\,kAU (V450~And~AB and V451~And).  

Once we have a detailed six-dimensional picture of the solar nighbourhood,
($x, y, z, \dot{x}, \dot{y}, \dot{z}$), possibly with the European Space
Agency {\em GAIA} mission, we will be able to corroborate or refute the
existence of ``ultra-wide'' binaries with separations over one~parsec,
such as $\alpha$~Lib and KU~Lib.
In the meantime, its existence is {\em not} a challenge for star formation
scenarios or studies of gravitational perturbations in the Galactic disc:
$\alpha$~Lib and KU~Lib simply constitute the most extreme example of young
wide binaries caught in the process of disruption by the Galactic tidal force.

%
%
%

\begin{acknowledgements}

I thank the anonymous referee for his/her helpful report and careful
reading of the manuscript, J. L\'opez-Santiago, D. Montes, and A. Klutsch
for their valuable assistance in the preparation of the list of star candidates
in the Castor moving group, and R.-D. Scholz and M. C. Smith for their
information on young field brown dwarfs and very wide halo binaries. 
Formerly, I~was an investigador Juan de la Cierva at the Universidad
Complutense de Madrid; 
currently, I~am an investigador Ram\'on y Cajal at the Centro de
Astrobiolog\'{\i}a.
This research has made use of the Washington Double Star Catalog maintained at
the United States Naval Observatory, the SIMBAD, operated at Centre de Donn\'ees
astronomiques de Strasbourg, France, and NASA's Astrophysics Data System.
Financial support was provided by the Universidad Complutense de Madrid,
the Comunidad Aut\'onoma de Madrid, the Spanish Ministerio Educaci\'on y
Ciencia, and the European Social Fund under grants:
AyA2008-06423-C03-03, 			
AyA2008-00695,				
PRICIT S-0505/ESP-0237,			
and CSD2006-0070. 			

\end{acknowledgements}

\appendix

\section{Online material}

\begin{figure*}
\centering
\includegraphics[width=0.99\textwidth]{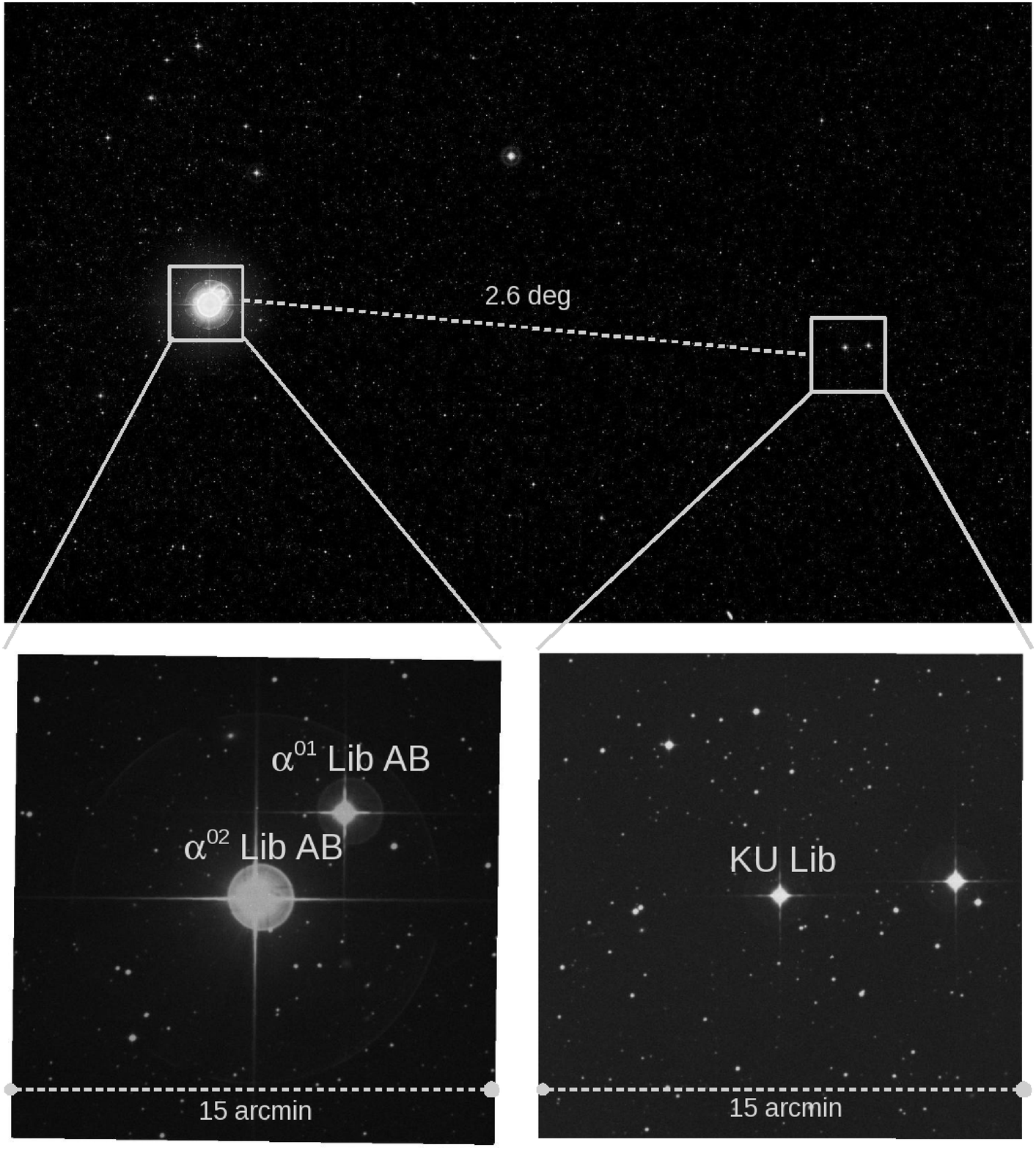}
\caption{{\em Top panel:} finding chart, about 3\,deg $\times$ 2\,deg,
showing the bright multiple system $\alpha^{02}$~Lib~AB and $\alpha^{01}$~Lib~AB
to the east (left) and KU~Lib to the west (right).
{\em Bottom panels:}  15\,arcmin $\times$ 15\,arcmin captions of the innermost
areas surrounding the stars from the SuperCOSMOS digitisations of the United
Kingdom Schmidt Telescope photographic plates. 
Some angular separations and fields of view are indicated.}  
\label{theimage}
\end{figure*}
%

\begin{longtable}{l cc c l}
\caption{\label{table.castor}{A compilation of star candidates in the Castor moving group.}}\\ %
\hline
            \noalign{\smallskip}
Name  		& $\alpha^{J2000}$	& $\delta^{J2000}$	& $d$ 		& Multiplicity 	\\
  		& 			& 			& [pc]		&  		\\
            \noalign{\smallskip}
\hline
            \noalign{\smallskip}
\endfirsthead
\caption{A compilation of star candidates in the Castor moving group (cont.).}\\
\hline
\hline
            \noalign{\smallskip}
Name  		& $\alpha^{J2000}$	& $\delta^{J2000}$	& $d$ 		& Multiplicity 	\\
  		& 			& 			& [pc]		&  		\\
            \noalign{\smallskip}
\hline
            \noalign{\smallskip}
\endhead
\hline
\endfoot
$\kappa$ Phe	& 00 26 12.20		& --43 40 47.4		& 23.81$\pm$0.09	& ... 		\\ %
QT And		& 00 41 17.34		&  +34 25 16.9		& 50$\pm$10 		& ... 		\\ 
EW Cet		& 01 16 24.20		& --12 05 49.2		& 25.9$\pm$0.5 		& ... 		\\ %
FN Cet		& 02 04 59.33		& --15 40 41.2		& 25.6$\pm$0.6 		& ... 		\\ %
V450 And AB	& 02 12 55.01		&  +40 40 06.0		& 26.8$\pm$0.4 		& SB, wide 	\\ 
V451 And	& 02 13 13.34		&  +40 30 27.3		& 26.2$\pm$0.4 		& Wide 		\\ %
LP 944--20	& 03 39 35.22  		& --35 25 44.1		& 4.97$\pm$0.10		& ... 		\\ %
HD 24053 AB	& 03 50 08.89 		&  +06 37 14.5		& 33.0$\pm$1.4 		& SB 		\\ %
HD 30957 AB	& 04 56 26.05 		&  +64 24 09.6		& 39.1$\pm$1.4 		& SB 		\\ %
V998 Ori AB	& 05 32 14.66 		&  +09 49 14.9		& 12.8$\pm$0.6 		& SB 		\\ %
HD 37216	& 05 39 52.35		&  +52 53 51.0		& 27.9$\pm$0.7 		& ... 		\\ %
$\zeta$ Lep	& 05 46 57.34		& --14 49 19.0		& 21.61$\pm$0.07 	& ... 		\\ %
V575 Pup AB	& 06 04 46.68 		& --48 27 29.9 		& 30.0$\pm$0.5 		& Close 	\\ %
HD 41842	& 06 06 16.61  		& --27 54 21.0		& 32.1$\pm$1.0 		& ..		\\ %
GJ 226.2	& 06 07 55.25  		&  +67 58 36.5		& 24.5$\pm$1.1 		& ..		\\ %
V356 CMa AB	& 06 39 11.63  		& --26 34 18.8		& 52$\pm$4 		& Close 	\\ %
HD 50255 AB	& 06 52 02.38 		& --11 12 16.2 		& 29.8$\pm$1.7 		& SB 		\\ %
HD 51825 AB	& 06 57 17.58  		& --35 30 25.8		& 43.1$\pm$0.8 		& Close 	\\ %
BL Lyn		& 07 31 57.33 		&  +36 13 47.4		& 12.0$\pm$0.6		& Wide		\\ %
VV Lyn AB	& 07 31 57.72 		&  +36 13 09.8		& 11.9$\pm$0.5		& SB, wide 	\\ 
Castor AB	& 07 34 35.86  		&  +31 53 17.8		& 15.6$\pm$0.9 		& Close 	\\ %
YY Gem		& 07 34 37.41  		&  +31 52 09.8		& 15.6$\pm$0.9 		& Wide 		\\ %
DX Cnc		& 08 29 49.34  		&  +26 46 33.7		& 3.63$\pm$0.04		& ... 		\\ %
CU Cnc AB	& 08 31 37.60  		&  +19 23 39.6		& 11.1$\pm$1.0 		& SB, wide 	\\ 
CV Cnc AB	& 08 31 37.44  		&  +19 23 49.5		& 11.1$\pm$1.0 		& SB, wide 	\\ 
HL Cnc		& 09 01 22.78 		&  +10 43 58.5		& 64$\pm$4		& ... 		\\ %
V405 Hya	& 09 04 20.69 		& --15 54 51.3		& 28.3$\pm$0.6		& ... 		\\ %
HD 79555 AB	& 09 14 53.66 		&  +04 26 34.4		& 18.0$\pm$0.5 		& SB 		\\ %
$\psi$ Vel AB	& 09 30 42.00  		& --40 28 00.4		& 18.81$\pm$0.13  	& Close 	\\ %
AD Leo		& 10 19 36.28  		&  +19 52 12.1		& 4.69$\pm$0.09 	& ... 		\\ 
36 UMa BC	& 10 30 25.31 		&  +55 59 56.8		& 12.80$\pm$0.05 	& SB 		\\ %
36 UMa A	& 10 30 37.58 		&  +55 58 49.9		& 12.80$\pm$0.05 	& Wide		\\ 
HD 93915 AB	& 10 51 14.62 		&  +46 47 46.6		& 42.6$\pm$1.6 		& SB 		\\ %
GY Leo		& 10 56 30.80  		&  +07 23 18.5		& 17.3$\pm$0.3 		& ... 		\\ %
XZ LMi AB	& 10 59 48.28 		&  +25 17 23.5		& 35.8$\pm$1.0 		& SB 		\\ %
Ross 104	& 11 00 04.26 		&  +22 49 58.7		& 6.66$\pm$0.08 	& ... 		\\ %
PR Vir		& 11 56 41.18 		& --02 46 44.2		& 42$\pm$2 		& ... 		\\ %
HD 119124 A	& 13 40 23.23  		&  +50 31 09.9		& 25.3$\pm$0.3 		& Wide	 	\\ %
HD 119124 B	& 13 40 24.51  		&  +50 30 57.6		& 25.3$\pm$0.3 		& Wide 		\\ %
GY Boo		& 14 12 41.56 		&  +23 48 51.5		& 33.3$\pm$1.2 		& ... 		\\ %
KU Lib		& 14 40 31.11 		& --16 12 33.4		& 23.7$\pm$0.3		& Wide 		\\ %
$\alpha^{01}$ Lib AB& 14 50 41.18 	& --15 59 50.1		& 23.0$\pm$0.2 		& Close, wide 	\\ %
$\alpha^{02}$ Lib AB& 14 50 52.71 	& --16 02 30.4		& 23.24$\pm$0.10 	& SB, wide 	\\ %
EV Dra AB	& 16 01 47.46 		&  +51 20 52.0		& 57$\pm$2		& SB 		\\ %
$\mu$ Dra AB	& 17 05 20.12 		&  +54 28 12.2		& 27.4$\pm$0.3 		& Close 	\\ %
$\mu$ Dra C	& 17 05 20.27 		&  +54 27 59.8		& 27.4$\pm$0.3		& Wide		\\ 
BD--46 11540	& 17 28 39.95 		& --46 53 42.7		& 4.54$\pm$0.03 	& ... 		\\ 
MS Dra		& 17 39 55.69 		&  +65 00 05.9		& 26.3$\pm$0.4 		& ... 		\\ %
BD+21 3245	& 17 53 29.94 		&  +21 19 31.0		& 24.4$\pm$0.6 		& ... 		\\ %
HD 168442	& 18 19 50.84 		& --01 56 19.0		& 19.6$\pm$0.6 		& ... 		\\ %
Vega		& 18 36 56.34 		&  +38 47 01.3		& 7.68$\pm$0.02 	& ... 		\\ %
V1436 Aql A	& 18 54 53.66 		&  +10 58 40.2		& 16$\pm$2	 	& Wide 		\\ %
V1436 Aql B	& 18 54 53.81 		&  +10 58 43.5		& 16$\pm$2	 	& Wide 		\\ %
V1285 Aql AB	& 18 55 27.41 		&  +08 24 09.0		& 11.8$\pm$0.2 		& SB 		\\ %
HD 181321	& 19 21 29.76 		& --34 59 00.6		& 18.8$\pm$0.5 		& ... 		\\ %
HD 186922 AB	& 19 39 06.37 		&  +76 25 19.3		& 29.7$\pm$0.5 		& Close 	\\ %
HD 191285	& 20 09 36.47 		& --14 17 12.9		& 31.6$\pm$1.4 		& ... 		\\ %
Alderamin	& 21 18 34.77 		&  +62 35 08.1		& 15.04$\pm$0.02 	& ... 		\\ %
G 263--10	& 21 58 24.52 		&  +75 35 20.6		& 20.8$\pm$0.5		& ... 		\\ %
V374 Peg	& 22 01 13.12 		&  +28 18 24.9		& 8.9$\pm$0.2		& ... 		\\ %
V447 Lac A	& 22 15 54.14 		&  +54 40 22.4		& 21.5$\pm$0.2		& Wide 		\\ %
V447 Lac B	& 22 16 02.59 		&  +54 39 59.5		& 21.5$\pm$0.2		& Wide		\\ 
TW PsA		& 22 56 24.05 		& --31 33 56.0		& 7.61$\pm$0.04		& Wide 		\\ %
Fomalhaut	& 22 57 39.05		& --29 37 20.0		& 7.70$\pm$0.03		& Wide		\\ 
HD 217107	& 22 58 15.54 		& --02 23 43.4		& 19.86$\pm$0.15	& ...		\\ 
HK Aqr		& 23 08 19.55 		& --15 24 35.8		& 22.3$\pm$1.1		& ... 		\\ %
KZ And AB	& 23 09 57.36 		&  +47 57 30.1		& 24$\pm$2		& SB, wide 	\\ %
HD 218739 AB	& 23 09 58.87 		&  +47 57 33.9		& 25.0$\pm$1.4		& SB, wide 	\\ %
NX Aqr		& 23 24 06.34 		& --07 33 02.7		& 30.1$\pm$0.6		& ... 		\\ %
EQ Peg AB	& 23 31 52.18 		&  +19 56 14.1		& 6.18$\pm$0.06		& Close 	\\ %
           \noalign{\smallskip}
\end{longtable}

\end{document}